**Probing the pathway of an ultrafast structural phase transition to illuminate the transition mechanism in $Cu_2S$**

Junjie Li[1], Kai Sun[2], Jun Li[1], Qingping Meng[1], Xuewen Fu[1], Wei-Guo Yin[1], Deyu Lu[3], Yan Li[4], Marcus Babzien[5], Mikhail Fedurin[5], Christina Swinson[5], Robert Malone[5], Mark Palmer[5], Leanne Mathurin[6], Ryan Mason[6], Jingyi Chen[6], Robert M. Konik[1], Robert J. Cava[7], Yimei Zhu[1], Jing Tao[1, *]

[1]Condensed Matter Physics & Materials Science Department, Brookhaven National Laboratory, Upton, NY 11973, USA

[2]Department of Physics, University of Michigan, Ann Arbor, MI 48109, USA

[3]Center for Functional Nanomaterials, Brookhaven National Laboratory, Upton, NY 11973, USA

[4]American Physical Society, 1 Research Road, Ridge, NY 11961, USA

[5]Accelerator Test Facility, Brookhaven National Laboratory, Upton, NY 11973, USA

[6]Department of Chemistry and Biochemistry, University of Arkansas, Fayetteville, AR 72701, USA

[7]Department of Chemistry, Princeton University, Princeton, NJ 08544, USA

Abstract

Disentangling the primary order parameter from secondary order parameters in phase transitions is critical to the interpretation of the transition mechanisms in strongly correlated systems and quantum materials. Here we present a study of structural phase transition pathways in superionic $Cu_2S$ nanocrystals that exhibit intriguing properties. Utilizing ultrafast electron diffraction techniques sensitive in both momentum-space and the time-domain, we distinguish the dynamics of crystal symmetry breaking and lattice expansion in this system. We are able to follow the transient states along the transition pathway and so observe the dynamics of both the primary and secondary order parameters. Based on these observations we argue that the mechanism of the structural phase transition in $Cu_2S$ is dominated by the electron-phonon coupling. This mechanism advances the understanding from previous results where the focus was solely on dynamic observations of the lattice expansion.

Introduction

Spontaneous symmetry breaking in phase transitions plays a central role in the study of emergent properties in a variety of quantum materials including superconductors, multiferroics, and topological materials. As pointed out by Landau, a phase transition can be described by the corresponding order parameter, a quantity that characterizes the symmetry breaking pattern of the phase transition and that demonstrates singular behavior across the phase boundary. The order parameter utilized in Landau's theory is also known as primary order parameter. In addition to the primary order parameter, in a real material system singular behavior will be seen for other characteristics that are not directly associated with the symmetry breaking. These are associated with secondary order parameters. The secondary order parameters can also be used to identify the phase boundary and they usually break fewer (or no) symmetry elements in comparison with the primary one. Although primary and secondary order parameters may mark the same phase boundary, they play completely different roles in the phase transition [1, 2]. Specifically, the primary order parameter is the origin of singular behavior across the phase boundary, while singularities in secondary order parameters are subsidiary being induced by their coupling to their attendant primary order parameters.

In many correlated electron systems and quantum materials, the primary and secondary order parameters are intertwined, leading to considerable ambiguity in understanding materials' functionalities that are associated with phase transitions [3-7]. However, recent advancements [8-11] in time-resolved ultrafast observation method using pump-probe techniques offer the ability to track independently the dynamics of both. They are able, in general, to probe both the steady state associated with the phase transition as well as the pathways taken by a material to achieve this steady state. We will demonstrate this here in an ultrafast electron diffraction study of $Cu_2S$ nanocrystals.

$Cu_2S$ is known as a superionic conductor with liquid-like behavior of copper ions above the critical temperature, giving rise to applications in electrochemical, thermoelectric, and battery devices due to its unique electrochemical performance [13-18]. The structural phase transition occurring at a critical temperature close to ambient has been studied intensively in $Cu_2S$ [19-23]. The structural phase transition involves crystal symmetry breaking of the crystalline lattice and a "simultaneous" volume expansion, i.e., $Cu_2S$ is monoclinic (space group *$P2_1/c$*; hereafter called the "L-phase") at room temperature and hexagonal (space group *$P6_3/mmc$*; hereafter called the "H-phase") at temperatures above 375 K. The structural phase transition in $Cu_2S$ is likely to be a first-order transition, evidenced by multiple reports of hysteresis during the transition [13, 14, 24, 25].

Previous ultrafast x-ray absorption work on $Cu_2S$ nanocrystals reported a reversible switching behavior between the L-phase and the H-phase with a time-constant of 18 ps [12]. This study however was only able to distinguish the dynamics of the secondary order parameter, the rate of lattice expansion. Here we will add to this prior work. Due to the sensitivity in momentum/reciprocal space of ultrafast electron diffraction, we will be able to report measurements of the transient crystal symmetry breaking (the primary order parameter) and lattice expansion during the phase transition.

In ref. [12] it was suggested that the transition is not governed by electronic mechanisms but by the copper ion self-diffusion rate. However, we will argue that crystal symmetry breaking, i.e. the primary order parameter, plays a key role and is indicative of the underlying physics of the material. The focus on both order parameters brings forth a basic question for a wide range of materials: what are temporal scales governing the two order parameters during the onset of a structural transition? A simplified case illustrating this issue using a 1D assembly of atoms is found in Fig. 1, which shows that a material can take multiple pathways on transforming from one phase to another in parameter space. In this letter we will identify which of these pathways $Cu_2S$ takes after being pumped with pulsed laser energy and illuminate the transition mechanism.

Results and Discussions

$Cu_2S$ nanoplates were synthesized (see ref. [24]) that are on the order of 10 nm thick and 100 nm in lateral dimension, and were then deposited on a transmission electron microscope (TEM) grid with β-$Si_3N_4$ support membrane at the center. The *c* axis of both the L-phase and the H-phase is perpendicular to the top and bottom surface of the nanoplates [24]. TEM low-mag images show most of the $Cu_2S$ nanoplates lie flat on the $\beta$-$Si_3N_4$ support with their $c$ axes aligned along the electron beam direction (supplementary Figure S2).

The UED experimental settings are the same as in ref. [11] and also can be found in the Methods in the supplementary material. Because the electron probe in the UED experiment is about 200 μm in diameter, the UED patterns were obtained from the scattering from millions of $Cu_2S$ nanoparticles, making the results statistically representative. A typical UED pattern (80 ps after the laser pumping) is shown in supplementary Figure S3 with the integrated radial intensity profile.

As shown in Figure 2(a), a comparison of the UED patterns before and after the laser pump indicates that the peak center moves toward the direct beam, which is consistent with the lattice expansion both due to laser heating and the phase transition to the high temperature phase. More significantly, the peaks in the UED patterns become sharper after the laser pumping. This cannot arise from a laser heating effect because heating induces larger atomic vibrations that would affect the intensity in the background and broaden the diffraction peaks, the opposite of what we observed.

It is necessary to understand the formation of the diffraction peaks in the UED patterns during the phase transition. Each diffraction peak in the UED patterns consists of multiple reflections with close $q$ values in the reciprocal space, which is illustrated in Fig. 2(b) using electron diffraction (ED) patterns obtained from individual $Cu_2S$ nanoplates in transmission electron microscopy (TEM). Firstly, the diffraction peaks in the UED patterns have the average intensity from many individual $Cu_2S$ nanoplates with aligned $c$ axes. Thus the radial position of the diffraction peaks in the UED patterns can be indicated by dash circles in ED patterns in Fig.

2(b) for the (110) and (120) peaks (using the hexagonal notation). Secondly, due to the instrumental broadening [26], the UED patterns do not have the resolution to distinguish reflections with close $q$ values in reciprocal space, particularly for the L-phase.

To be more specific, for the hexagonal H-phase, the (110) reflections in the ED pattern have single $q$ value of 5.02 nm$^{-1}$ and, similarly, a single $q$ value of 7.67 nm$^{-1}$ is related to the (120) reflections at 325 ºC [21]. Thus the (110) and (120) peaks in the UED pattern in Fig. 2(a) are formed by those single-$q$-valued reflections convoluted with instrument broadening. On the other hand for the monoclinic L-phase, as highlighted by open-green squares (Fig. 2(b)), the (110) and (120) reflections have multiple q-values due to its lower crystal symmetry. Not only do the single-$q$-valued (110) and (120) reflections in the H-phase split into multiple $q$ values, but additional reflections also appear (highlighted by open-green triangles in Fig. 2(b)), giving an additional broadening of the peak intensity profile. A qualitative construction of the intensity profiles of the UED peaks is described in Fig. 2(c) and 2(d) for the (110) and (120) peaks. The additional broadening of the UED peaks from the L-phase can be clearly seen from the intensity profiles. We therefore conclude that the observation of the sharpening of the UED peaks after laser pumping is clear evidence of the structural phase transition from L-phase to the H-phase in the Cu$_2$S nanoplates.

The peak width (reflecting the change of crystal symmetry) and peak center (reflecting the change in molar volume) change as a function of time delay are shown in Figure 3(a) and 3(b) for the (110) and (120) peaks, respectively. We note that, the changes of the peak centers for both (110) peak and (120) peak are ~ 0.8%, consistent with previous study reported for the L-phase at room temperature and the H-phase at 325 ºC [19, 21]. The temporal evolutions of peak center and peak width are obviously very distinct in the ultrafast-time domain. In particular, for the (110) peak in Figure 3(a), the peak width changes from ~ 0.44 nm$^{-1}$ to 0.36 nm$^{-1}$ with a time constant of 2.1 ± 0.2 ps using exponential fitting, whereas the time constant of peak center position is about 6.7 ± 0.4 ps. For the (110) peak in UED, the disappearance of the green-triangle-marked reflections (on one side of the (110) green-square reflections in Fig. 2(b)) affects both the peak width and peak center position. This explains the plot that the peak center measurements change synchronously with the width measurements in the first few ps after the pump. The observation of the (120) peak, shown in Fig. 3(b), delivers the same message, except that the separation of the temporal evolution of peak width and peak center position is clearer: the time constant is ~ 2.3 ± 0.2 ps for the peak width and ~ 10.2 ± 0.3 ps for the peak center. This even more clear distinction is attributed to the fact that the green-triangle-marked reflections (in Fig. 2(b)) of the L-phase are located symmetrically around the (120) reflections, and the disappearance of the reflections of the L-phase during the transition affects the peak width measurements but has much less effect on the peak center measurements. In other words, the difference between the time constants (6.7 ± 0.4 ps for the (110) peak and 10.2 ± 0.3 ps for the (120) peak) are resulted from the technique with instrument broadening and the characteristics of the crystal structure. Even so, the temporal measurements of the evolution of the (110) peak and (120) peak strongly indicate that the crystal symmetry breaking takes place much faster than the lattice expansion. Thus a pathway similar to the orange route in Fig. 1 is indicated by our ultrafast observations in this material. We also provide a theoretical discussion of the dynamics of both the primary and secondary order parameters, based on the theory developed in [1, 27], in the supplementary material.

    Other considerations that might affect the measurements here are taken into account. A number of previous studies reported that lattice expansion starts from the surface of a material

and propagates elastically through the material by the speed of sound, which can be calculated by the mass density and bulk modulus [28, 29]. Taking a bulk modulus ($B_T$) of 50.95 GPa and a mass density (ρ) of 5.6 g/cm$^3$ [15], the speed of sound $v = \sqrt{\frac{B_T}{\rho}}$ is approximately 3016 m/s in bulk Cu$_2$S. Since the dimension (thickness and the lateral size) of the Cu$_2$S nanoparticles varies from 10 nm to 100 nm, the time for the propagation of lattice expansion ranges from a few ps to tens of ps, falling into the level of our observation of peak center evolution in time. Thus the evolution of lattice expansion observed here may be a combination of the phase transition and elasticity of the material. Moreover, the shear modulus ($G_T$) that might affect the symmetry of the structure is significantly smaller than the bulk modulus in Cu$_2$S ($G_T$ = 17.77 GPa) [15]. It means that if the shear modulus could dominate the crystal symmetry change through the elasticity, the process would be strongly correlated with the establishment of mechanical equilibrium, and slower than the observed lattice expansion. That is not the case for Cu$_2$S in our measurement. Note that the elastic properties could also affect a pressure that is built up through deformation potential at a time scale as fast as phonons are absorbed [30-32] and could drive phase transition in other systems [33]. However, the details of the possible deformation potential in Cu$_2$S are beyond the scope of this work.

Our dynamic observations of the structural phase transition also settle a longstanding debate of the role of the electronic states in the phase transition of Cu$_2$S. In particular, based on the time constant (~ 18 ps) of lattice expansion during the structural phase transition using near-edge X-ray spectroscopy reported previously on Cu$_2$S, it was concluded that the structural phase transition takes place on a time scale that is much longer than the electronic excited states needs to respond to the pumping process; thus the role of the electronic states in the transition mechanism was considered negligible [12]. However, X-ray spectroscopy results come from the scattering beam with wave vector $q \approx 0$, so the measurements do not directly reflect the process of structural transition. The conclusion of Ref. [12] is inconsistent with other experimental results. For instance, in a recent TEM study, $Cu_2S$ was found to exhibit the surprising characteristic that purely electronic excitations can reversibly manipulate the crystal's structural transformation between its L- and H-phases [24]. Since the observations strongly imply an intimate interplay between electronic degrees of freedom and the lattice, it is difficult to imagine how the modification of the electronic states can be ruled out as the mechanism of the structural transition in this material. Nevertheless as TEM observations cannot trace the structural dynamics through the transition, they cannot pin down the driving force of the transition.

The above ambiguity concerning the phase transition in $Cu_2S$ and other electronic materials has to be addressed by direct probes of the structural evolution using momentum-resolved techniques in the ultrafast time domain. Our measurement of the temporal evolution of lattice expansion (~ 10 ps) is on the same order of the result from the X-ray spectroscopy work on $Cu_2S$ [12]; however, we find that the dynamics of crystal symmetry breaking (~ 2ps) is faster by 3 - 5 times. Indeed, the time scale of crystal symmetry breaking measured here coincides with the time scale of the carrier population or carrier excited states measured by optical pump–probe spectroscopy in this material [12], directly indicating a key role that electronic states play in the structural transition. In addition, it is accepted that, during a pump-probe process, electron-phonon scattering takes place from hundreds of femtoseconds to a few ps after the laser pumping [34-36]. Governed by electron-phonon coupling, the lattice symmetry changes from monoclinic to hexagonal but the unit-cell maintains the size of the L-phase in $Cu_2S$ for several ps. In the subsequent tens of ps, the lattice continuously expands to the value of the H-phase, resulting from the diffusion of the copper ions and elastic properties of the material, so completing the transition.

We believe our finding could inspire further study of the structural instability in many superionic conductors such as $Cu_2S$, CuBr, $Ag_2S$ and AgI by exploring how the perturbation of electronic states affects the bonding situation either locally or collectively. Those ionic conductors share a common feature in their cations having a +1 electronic valence state. In the $Cu_2S$ case, the $Cu^+$ ions have the $d^{10}$ electronic configuration and the electronic band structure at the Fermi level consists of the Cu 4$s$ and 3$d$ bands [37]. This points to two following ways for electron-phonon coupling to exert its role in the photoinduced phase transition:

1) A photoexcitation causes a $3d^{10}$-$3d^94s^1$ transition of the Cu ions. Since 4$s$ wide-band electrons are expected to move much faster than 3$d$ narrow-band holes, the lag of the latter yields effective $Cu^{2+}$ $3d^9$ transient states, which drive the breathing-mode and Jahn-Teller-mode motions of the surrounding sulphur anions.

2) Given the fact that there are noticeable Cu vacancies in the real material $Cu_{2-x}S$ [14-19], Cu ions already exist as $Cu^{2+}$ in small proportions and are likely randomly distributed. Hence, photoexcitation may cause electron transfer between neighboring $Cu^{1+}$ and $Cu^{2+}$ cations (i.e., $Cu^{1+} \rightarrow Cu^{2+}$ and $Cu^{2+} \rightarrow Cu^{1+}$), inducing a change in the lattice distortion surrounding the pair of $Cu^{1+}$ and $Cu^{2+}$ ions, in the same spirit as the photoinduced intersite $Mn^{3+}$-$Mn^{4+}$ transition in the manganites [11]. This facilitates the crystal lattice to become more disordered with higher crystal symmetry. These two electron-phonon coupling mechanisms are compatible with each other and can work together.

In summary, our UED-based time-resolved observations of the temporal evolution of the lattice emphasize that both the primary and secondary lattice parameters need to be monitored in order to come to a complete understanding of the Cu$_2$S phase transition. Our UED results suggest an intimate electron-lattice interplay, e.g., electron-phonon coupling, to be the mechanism by which the structural phase transition is mediated in this material. Moreover, knowing the atomic arrangements at each snapshot during the phase transition is essential for characterizing the evolution of lattice energy, making it possible to quantitatively depict the energy pathways of phase transitions for other materials in the future. The capability demonstrated in this study will provide guidelines for phase transitions and will significantly impact research in correlated electron systems and quantum materials.

Supplementary Material

Technical details of sample preparation, experimental setup, UED beam, diffraction analysis and a theoretical prediction of the dynamics of the order parameters are included with four supplementary figures and one supplementary table.

Acknowledgements

We thank Dr. Lijun Wu and Tatiana Konstantinova for their discussion of the work. Research was sponsored by DOE-BES Early Career Award Program and by DOE-BES under Contract DE-SC0012704. The work at University of Michigan was supported by NSF-EFMA-1741618 and the Alfred P. Sloan Foundation. D.L. was supported by the Center for Functional Nanomaterials, which is a U.S. DOE Office of Science Facility, at Brookhaven National Laboratory. J.C., L.M. and R.M. was supported in part by the University of Arkansas and the National Science Foundation (NSF) through Center for Advanced Surface Engineering under Grant no. OIA-1457888 and the Arkansas EPSCoR Program, ASSET III to J.C. The work at Princeton was supported by Department of Energy, Division of Basic Energy Sciences, Grant DE-FG02-98ER45706.

Author Contributions

J.T. and Junjie Li designed the project and were the primary interpreter of the resulting data. J.C., L.M. and R.M. synthesized nanomaterials. J.T., Junjie Li., Jun Li., M.B., M.F., C.S., R.M., M.P. and Y.Z. collected UED data. J.T., Junjie Li. and Jun Li. analyzed UED data. J.T., K.S., Q.M., W-G.Y., R.M.K., R.J.C. and Y.Z. performed the study of the transition mechanism. J.T. Y.L., D.L. and R.J.C. performed the study of crystal structures. All authors contributed to the writing of the paper.

Author Information

Correspondence and requests for materials should be addressed to J. T. (jtao@bnl.gov)

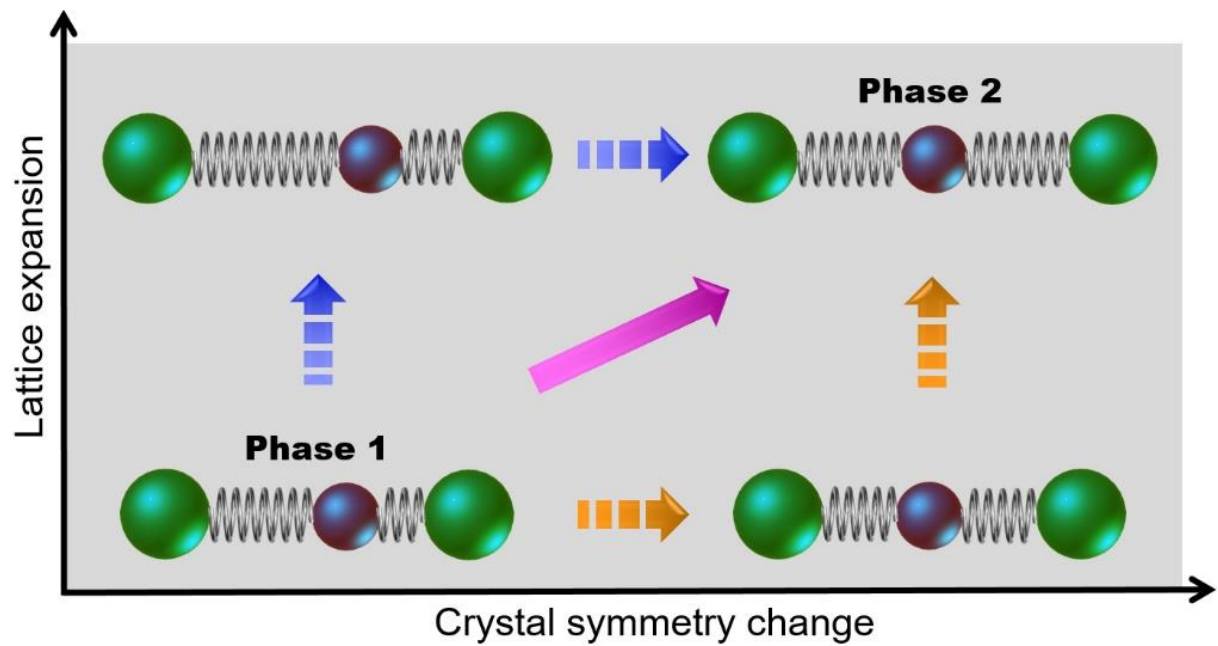

**Figure 1.** Schematic of a structural phase transition that involves both the crystal symmetry breaking and lattice expansion in a 1D assembly of atoms. Phase 1 has two types of atoms that have unequal spacing. Phase 2 has the same group of atoms with altered lattice symmetry and spacing. Arrows in orange, purple and blue indicate possible transition pathways in the ultrafast time domain.

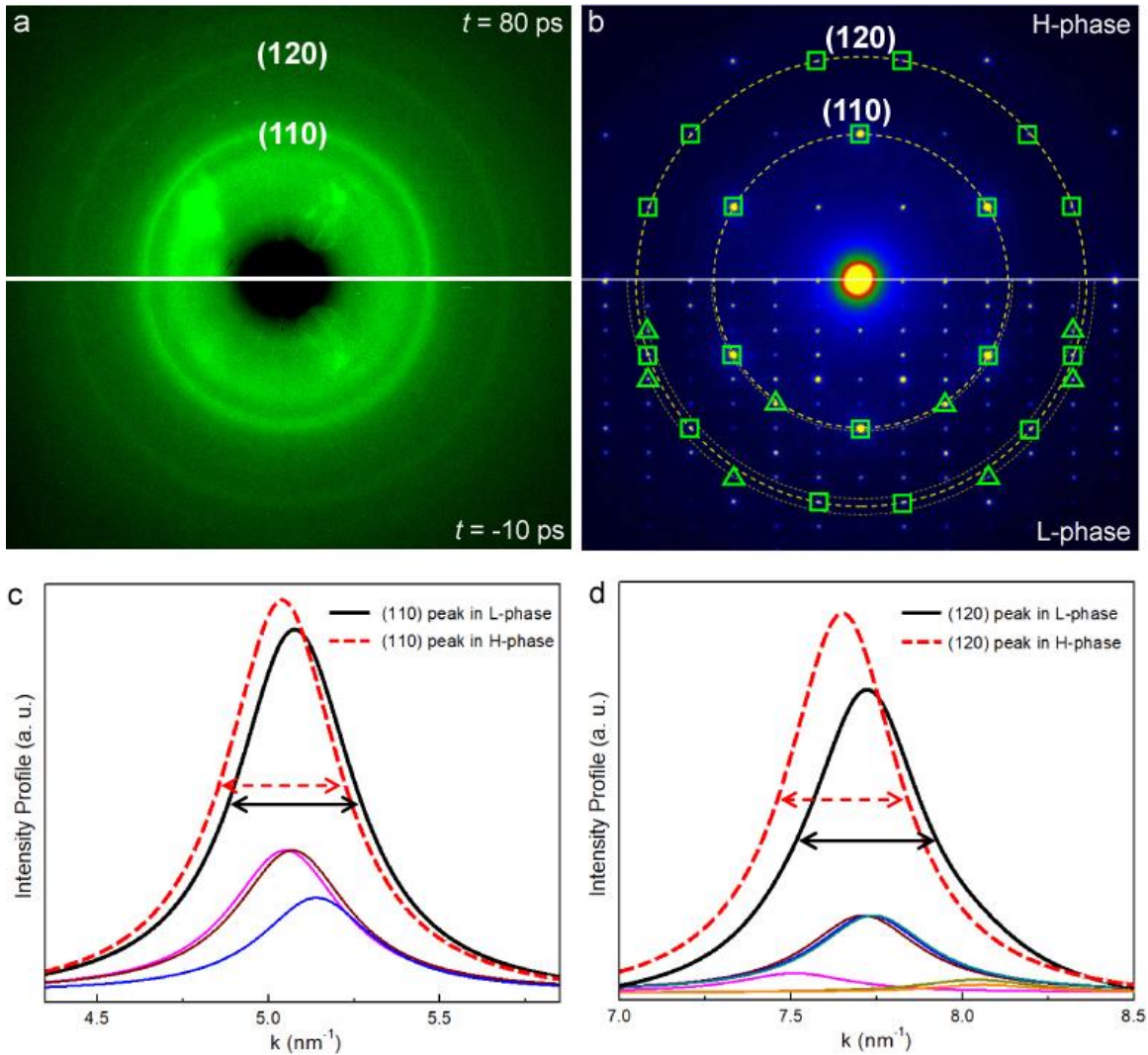

**Figure 2.** (a) A comparison of UED patterns taken before (bottom) and after (top) laser pump. (110) and (120) reflection rings are indexed using the notation of the hexagonal structure. UED patterns were recorded over an integration of 200 shots, under a pump fluence of 2 mJ/cm$^2$. (b) Presented are the electron diffraction (ED) patterns obtained in the L-phase (bottom) and H-phase (top) in transmission electron microscopy (TEM). The open squares mark diffraction peaks that appear in both the L- and H-phases while the open triangles mark peaks that only appear in the L-phase. The UED rings observed in top part of (a) (after the pump) are formed from single diffraction peaks appearing in the TEM image of (b), namely the 5.02 nm$^{-1}$ peak for the (110) ring and the 7.67 nm$^{-1}$ peak for the (120) ring. In contract the rings in the UED pattern

in the bottom part of (a) (before the pump) are formed from reflections at a variety of $q$ values. The (110) ring is formed from reflections at 5.05 nm$^{-1}$, 5.07 nm$^{-1}$, and 5.13 nm$^{-1}$ while the (120) ring is composed primarily of reflections at 7.71 nm$^{-1}$, 7.73 nm$^{-1}$, and 7.74 nm$^{-1}$. This latter ring also sees contributions from reflections close by which however cannot be separately resolved because of instrument resolution. (c) The center and width of the (110) ring before (L-phase) and after (H-phase) the pump are schematically constructed. The black curve marks the ring before the pump when the material is in the L-phase. It is formed from a number of smaller peaks whose intensity profiles are given by the magenta, brown, and blue curves. The red curve gives the peak intensity profile for the H-phase – because of the higher symmetry of the hexagonal phase, this intensity profile is formed from a single ring as explained previously. (d) A similar construction for the intensity profile of the (120) ring is made. Here the intensity profile for the L-phase is composed of four separate intensity profiles (marked by the brown, blue, cyan, magenta, green and orange curves). Details of all these peaks can be found in supplementary Figure S4 and Table S1.

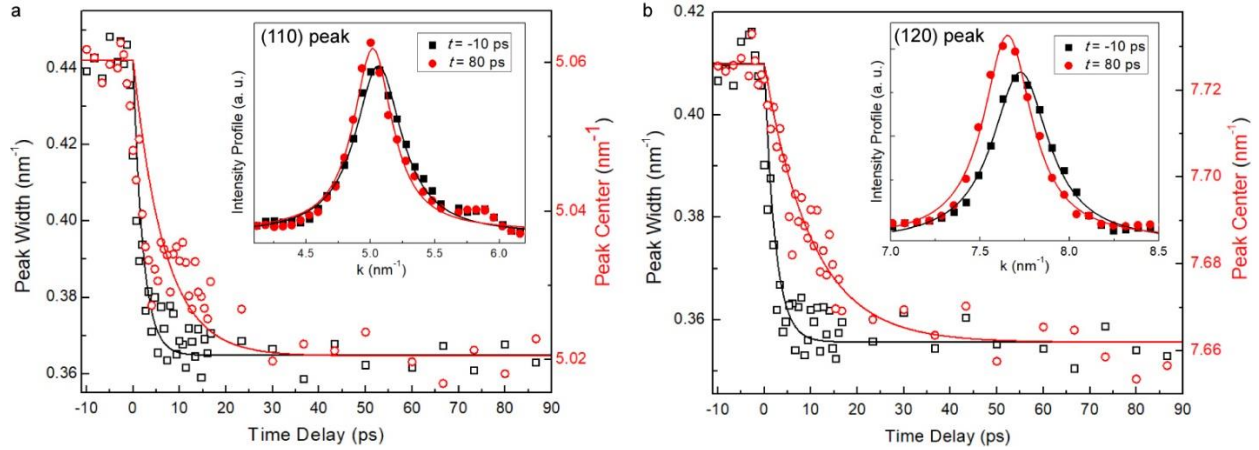

**Figure 3.** The change of peak width and center position after laser pump are presented. Peak width (plotted by black squares) and peak center position (plotted by red circles) in reciprocal space are given as a function of time delay for the (110) peak in (a) and the (120) peak in (b). The time constant of peak width change is ~ 2.1 ± 0.2 ps for the (110) peak and ~ 2.3 ± 0.2 ps for the (120) peak, whereas the time constant of the shift in the peak center position is ~ 6.7 ± 0.4 ps for the (110) peak and ~ 10.2 ± 0.3 ps for the (120) peak. Black and red lines are guide for eyes. Peak width was measured using the full width of half maximum of the peak intensity profiles, which are shown as the insets of the plots with a Lorentzian curve fitting. A hump in the shoulder of the (110) peak in the inset of a) is the (200) peak. We checked that this latter peak has a minimal effect on the extraction of the time constants from the presented measurements.

**Supplementary Material**

**Probing the pathway of an ultrafast structural phase transition to illuminate the transition mechanism in Cu$_2$S**

Junjie Li[1], Kai Sun[2], Jun Li[1], Qingping Meng[1], Xuewen Fu[1], Weiguo Yin[1], Deyu Lu[3], Yan Li[4], Marcus Babzien[5], Mikhail Fedurin[5], Christina Swinson[5], Robert Malone[5], Mark Palmer[5], Leanne Mathurin[6], Ryan Mason[6], Jingyi Chen[6], Robert M. Konik[1], Robert J. Cava[7], Yimei Zhu[1], Jing Tao[1, *]

[1]Condensed Matter Physics & Materials Science Department, Brookhaven National Laboratory, Upton, NY 11973, USA

[2]Department of Physics, University of Michigan, Ann Arbor, MI 48109, USA

[3]Center for Functional Nanomaterials, Brookhaven National Laboratory, Upton, NY 11973, USA

[4]American Physical Society, 1 Research Road, Ridge, NY 11961, USA

[5]Accelerator Test Facility, Brookhaven National Laboratory, Upton, NY 11973, USA

[6]Department of Chemistry and Biochemistry, University of Arkansas, Fayetteville, AR 72701, USA

[7]Department of Chemistry, Princeton University, Princeton, NJ 08544, USA

1.   **Methods**

Sample and experimental setup

$Cu_2S$ nanocrystalline plates were synthesized using the method described in ref. 24. A few drops of the corresponding $Cu_2S$ colloid were then deposited on a 15-nm-thick $\beta$-$Si_3N_4$ membrane TEM grid. After the solution was allowed to evaporate, the sample was characterized at room temperature using a 200 keV double Cs-corrected transmission electron microscope (JEOL ARM 200F). The diffraction patterns so obtained clearly show the different structures at room temperature and high temperatures, as shown in Fig. 2(b). Low-mag TEM images were also taken to examine the morphology of the spread $Cu_2S$ nanoplates on the $\beta$-$Si_3N_4$ membrane window ($0.25 \times 0.25$ mm$^2$ in lateral size). A good coverage of the nanoplates that do not colligated heavily can yield electron diffraction patterns with reasonable intensity/noise ratio in the MeV-UED measurement. In addition, previous structural characterizations showed that $Cu_2S$ nanoplates undergo the same structural phase transition as the bulk, except the fact that the transition temperature of the nanoplates is about 20 K below that of the bulk materials [24]. Placed on the top of the amorphous membrane, the $Cu_2S$ nanoplates are considered to be nearly freestanding during the structural phase transitions.

UED beam

In our UED experiments, optical pulses with a duration of 100 fs and a center wavelength of 800 nm (1.55 eV) were focused down to 1.5 mm on the sample to trigger electronic excitations and crystal structure evolution. At a specific time delay, well-synchronized 2.8 MeV electron pulses with a time resolution of 130 fs were collimated to 200 µm in the pumped area. Each diffraction pattern was recorded over an integration of 200 shots, under a pump fluence of 2 mJ/cm$^2$, at various pump-probe time delays. High-quality electron beams ($10^6$ electrons per bunch, of length 100 fs, with a longitudinal and transverse coherence length of ~2 nm and ~10 nm, respectively) were produced by using a unique BNL-type photocathode RF gun. The ultrahigh electron energy significantly minimizes space-charge effects, allowing for a high flux of electrons in extremely short pulses.

Diffraction analysis

For each UED pattern at various time delays, the center of the UED pattern is first determined according to the symmetry of diffraction rings. Then the radial intensity profile of diffraction patterns are plotted as a function of the distance to the center of the pattern, shown in supporting Fig. S3. Individual peaks of (110) and (120) are fitted with a combined Lorentzian function to determine the peak center position and peak width. Only the area in the UED pattern that is not polluted by the dark current is used for intensity analysis and measurements. The direct beam is blocked during the experiment and the intensities at the wave numbers of 4 nm$^{-1}$ and below are dominated by the reflections from the $\beta$-$Si_3N_4$ supporting membrane.

2.   **Theoretical prediction of the dynamics of the order parameters**

We compute the time scales of the (primary and secondary) order parameter decay after the material driven from a low-temperature phase to a high-temperature phase by an ultrafast laser pump.

Summary of results:

(1) In the short-time limit, the secondary order parameter varies significantly slower than the primary one ($t^2$ from low T to high T, and $t^3$ from high T to low T)
(2) pump from low T to high T: the variation of the secondary order parameter cannot be faster than the square of the primary one (See the inequality below $\frac{\eta(t)}{\eta_0} \geq \frac{Q(t)^2}{Q_0^2}$)
(3) pump from high T to low T, the variation of the secondary order parameter is no faster than the primary one

## 1a)    Second order phase transition

Here we start by considering a second order phase transition. However, as will be shown below, the main conclusions can be generalized to other cases, such as first-order phase transitions.

Considering a second order phase transition with a primary order parameter $Q$ and a second-order parameter $\eta$. The Ginzburg–Landau free energy is given by R. A. Cowley in ref [1].

$$F = \frac{a}{2}(T - T_c)Q^2 + u\, Q^4 + e\, Q^2\, \eta + \frac{1}{2}C\, \eta^2$$

By minimizing the Ginzburg–Landau free energy, it is straightforward to show that this system contains a second-order phase transition at $T_c$. In the high-temperature phase ($T > T_c$), $Q = \eta = 0$, while in the low-temperature phase ($T < T_c$), $Q = \sqrt{\frac{C\, a\, (T_c - T)}{4C\, u - 2e^2}}$ and $\eta = \frac{e\, a\, (T - T_c)}{4C\, u - 2e^2}$. Notice that here $Q$ scales $\sqrt{T_c - T}$, while $\eta \propto T - T_c$. This different scaling behavior reflects the fact that the scaling exponents of a secondary order parameter is two times larger than the primary one, as mentioned in the main text.

To model a phase transition from the low-$T$ to high-$T$ phase driven by ultra-fast laser pump, we set the control parameter $T$ as a function of time

$$T(t) = \begin{cases} T_i & t < 0 \\ T_f & t > 0 \end{cases}$$

with $T_i < T_c$ and $T_f > T_c$. Notice that here, $T(t)$ is a control parameter in this phenomenological description, and is not the equilibrium temperature for $t > 0$ where the system is far away from equilibrium.

To compute the real-time dynamics of the order parameters, we employ the Landau-Khalatnikov formalism [26], where that order parameters are assumed to be overdamped and thus terms of the form $\partial_t^2 Q$ and $\partial_t^2 \eta$ are ignored in the equations of motion. For $t > 0$, the dynamics of the order parameters are governed by the following equations

$$-\gamma_Q \partial_t Q = \frac{\partial F}{\partial Q} = a(T_f - T_c)Q + 4u\, Q^3 + 2\, e\, Q\, \eta$$

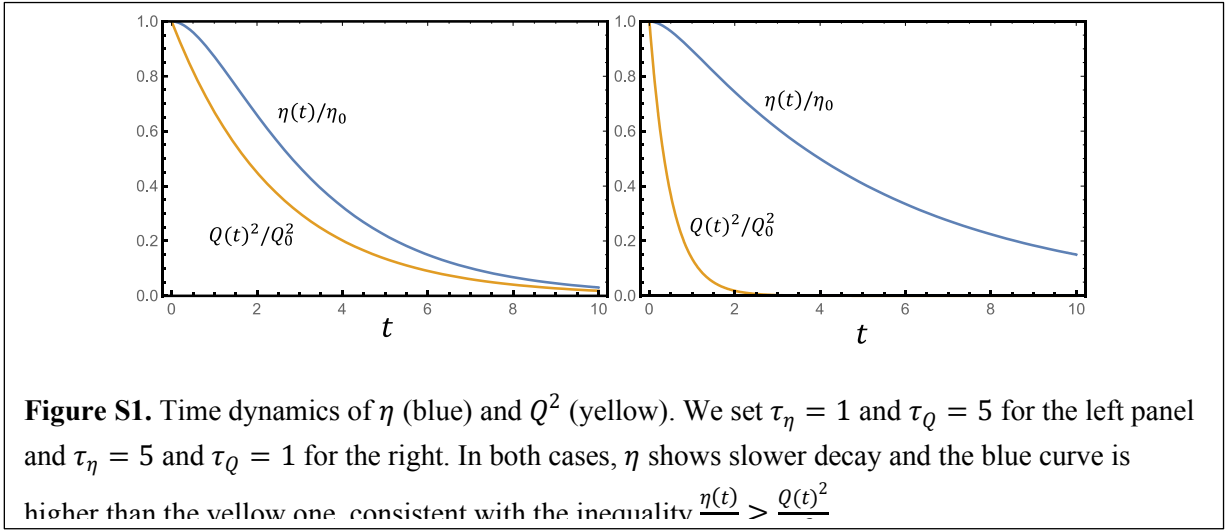

**Figure S1.** Time dynamics of $\eta$ (blue) and $Q^2$ (yellow). We set $\tau_\eta = 1$ and $\tau_Q = 5$ for the left panel and $\tau_\eta = 5$ and $\tau_Q = 1$ for the right. In both cases, $\eta$ shows slower decay and the blue curve is higher than the yellow one, consistent with the inequality $\frac{\eta(t)}{\eta_0} > \frac{Q(t)^2}{Q_0^2}$

$$-\gamma_\eta \partial_t \eta = \frac{\partial F}{\partial \eta} = e\, Q^2 + C\, \eta$$

Here $\gamma_Q$ and $\gamma_\eta$ are appropriate (phenomenological) coefficients of viscosity. And we utilize the initial condition $Q(t=0) = Q_0$ and $\eta(t=0) = \eta_0$, where $Q_0 = \sqrt{\frac{C\, a\, (T_c - T_i)}{4C\, u - 2e^2}}$ and $\eta_0 = \frac{e\, a\, (T_i - T_c)}{4C\, u - 2e^2}$ are the equilibrium expectation values of the order parameters at $t < 0$.

To simplify the calculation, here we take the leading order approximation and ignore all higher order terms in the equation of motion of $Q$.

$$-\gamma_Q \partial_t Q = a(T_f - T_c)Q$$

As for the equation of motion for $\eta$,

$$-\gamma_\eta \partial_t \eta = e\, Q^2 + C\, \eta$$

we keep both the linear term of $\eta$ and the quadratic terms of $Q$. As shown above, the scaling exponent of the secondary order parameter is twice of the primary one, and thus the $Q^2$ and $\eta$ terms are of the same order.

With this approximation, the two equations can be solved analytically.

$$Q(t) = Q_0 \exp(-\frac{t}{\tau_Q})$$

$$\eta(t) = \eta_0 \frac{\frac{\tau_Q}{2}\exp(-\frac{t}{\tau_Q/2}) - \tau_\eta \exp(-\frac{t}{\tau_\eta})}{\frac{\tau_Q}{2} - \tau_\eta}$$

Here, $\tau_Q = \frac{\gamma_Q}{a\,(T_f-T_c)}$ and $\tau_\eta = \frac{\gamma_\eta}{C}$. The primary order parameter decays exponentially, while the secondary order parameter has a more complicate functional form. From the solution above, it is straightforward to prove the following inequality:

$$\frac{\eta(t)}{\eta_0} \geq \frac{Q(t)^2}{Q_0^2}$$

As shown in Fig. S1, this inequality implies that *the secondary order parameter can never decay faster than the square of the primary order parameter*. To prove this inequality, one can compute $\frac{\eta(t)}{\eta_0} - \frac{Q(t)^2}{Q_0^2} = \tau_\eta \frac{\exp(-\frac{t}{\tau_Q/2}) - \exp(-\frac{t}{\tau_\eta})}{\frac{\tau_Q}{2} - \tau_\eta}$ and then observe that the l.h.s. of this equation is a non-negative function.

**1b)    The short-time limit**

In the short-time limit (i.e., $t$ is positive and small), $Q(t)$ and $\eta(t)$ take the following asymptotic form

$$Q(t) = Q_0 - Q_0 \frac{t}{\tau_Q} + O(t^2)$$

$$\eta(t) = \eta_0 - \eta_0 \frac{t^2}{\tau_Q \tau_\eta} + O(t^3)$$

In this limit, the secondary order parameter varies significantly slower than the primary one. As shown in the asymptotic form, the change of $\eta$ is $\sim t^2$, while the primary order parameter's change is linear in $t$.

This conclusion and the asymptotic form are generic, and they remain valid even if more complicated terms are added to the free energy, because of the following reason. At $t < 0$, thermal equilibrium requires $\frac{\partial F}{\partial Q} = \frac{\partial F}{\partial \eta} = 0$. At time $t = 0$, as we suddenly switch the coefficient of the $Q^2$ term, $\frac{\partial F}{\partial Q}$ immediately becomes nonzero, but $\frac{\partial F}{\partial \eta}$ still remains 0 (Notice that the $Q^2$ term doesn't arise in $\frac{\partial F}{\partial \eta}$ and thus a sudden change of its coefficient has no effect on $\frac{\partial F}{\partial \eta}$ at very small $t$). This is the fundamental reason why linear $t$ term is absent in the asymptotic of $\eta$.

### 1c) The long-time limit

In the long-time limit ($t \to +\infty$), the asymptotic form of $\eta(t)$ is

$$\eta(t) = \begin{cases} \eta_0 \exp(-\frac{t}{\tau_\eta}) & \tau_\eta > \tau_Q/2 \\ \eta_0 \exp(-\frac{t}{\tau_Q/2}) & \tau_\eta < \tau_Q/2 \end{cases}$$

If $\tau_\eta > \tau_Q/2$, the decay time for $\eta$ is set by $\tau_\eta$. For $\tau_\eta < \tau_Q/2$, the decay time of $\eta$ is locked to half the decay time of $Q$, which is also the decay time for $Q^2$, $Q(t)^2 = Q_0^2 \exp(-\frac{t}{\tau_Q/2})$. In both cases, the decay of $\eta$ cannot be faster than that of $Q^2$.

Why does $Q^2$ sets the upper bound for the decay speed of $\eta$? This question can be answered if we consider the $\tau_\eta = 0$ limit. In this limit, $\eta$ has infinite fast dynamics and thus it sets the upper limit for the decay rate. From the equation of motion above, at $\tau_\eta = 0$, $\eta(t) = -e\, Q(t)^2/C$.

Here the value of $\eta$ is locked to $\propto Q^2$, and thus these two quantities must have the same decay rate.

### 1d) First-order phase transition

For a first-order phase transition, the time dynamics of the primary order parameter can no longer be calculated using the Landau-Khalatnikov formulism above, because the time evolution of the primary order parameter here is more complicated and involves tunneling from one quasi-stable state to another stable state. However, as long as the time dependence of the primary order parameter takes an exponential-decay form, we can still utilize the same Landau-Khalatnikov approached to compute the dynamics of the secondary order parameter.

Consider the following Ginzburg–Landau free energy

$$F = F_Q + e\, Q^2\, \eta + \frac{1}{2} C\, \eta^2$$

where $F_Q$ is the free energy for $Q$. Regardless of details, here we assume that after ultra-fast laser pump, the primary order parameter start to decays exponentially to zero with decay time $\tau_Q$

$$Q(t) = Q_0 \exp(-\frac{t}{\tau_Q})$$

For $\eta$, we use the Landau-Khalatnikov approach,

$$-\gamma_\eta \partial_t \eta = \frac{\partial F}{\partial \eta} = e\, Q^2 + C\, \eta$$

As far as $\eta$ is concerned, it has the same initial condition and the same equation of motion, and thus all the conclusions obtained above shall remain.

### 1e) Higher order couplings between $Q$ and $\eta$

In addition to the $Q^2\, \eta$ term, higher order terms like $Q^{2n}\, \eta$ is allowed by symmetry in the Ginzburg–Landau free energy, where $n$ is an integer. These terms may become non-negligible for a first order phase transition, when the change of $Q$ is not small across the phase boundary. These higher order terms will not change the structure of the short-time and long-time asymptotic forms of $\eta$. For the short-time regime, these higher order terms will not change the fact that $\frac{\partial F}{\partial \eta} = 0$ immediately after $t = 0$, and thus the decay of $\eta$ at small $t$ cannot contain a linear $t$ term. In the long-time limit, as $Q$ becomes small (as a result of the exponential decay), higher order terms $Q^{2n} \eta$ with $n > 1$ becomes negligible. Thus, we will only need to keep the $Q^2 \eta$ term, which recovers with the case studied above.

## 1f) From the high-temperature phase to the low temperature phase

Here we consider the situation where a system in the high-temperature phase at $t < 0$ is suddenly pumped into a low-temperature phase at $t = 0$. For the primary order parameter, we assume that the order parameter evolves as

$$Q(t) = Q_0 \left[1 - \exp(-\frac{t}{\tau_Q})\right]$$

where $Q_0$ is the low-temperature expectation value, and then we compute the time dynamics of the secondary order parameter based on the Landau-Khalatnikov formulism

$$-\gamma_\eta \partial_t \eta = \frac{\partial F}{\partial \eta} = e\, Q^2 + C\, \eta$$

with initial condition $\eta(t = 0) = 0$.

The solution of the partial differential equation is

$$\eta(t) = \eta_0 \left[1 - \frac{\tau_\eta^2}{(\tau_Q - \tau_\eta)\left(\frac{\tau_Q}{2} - \tau_\eta\right)} e^{-\frac{t}{\tau_\eta}} - \frac{2\tau_Q}{\tau_Q - \tau_\eta} e^{-\frac{t}{\tau_Q}} + \frac{\tau_Q}{\tau_Q - 2\tau_\eta} e^{-\frac{2t}{\tau_Q}}\right]$$

where $\tau_\eta = \frac{\gamma_\eta}{C}$ and $\eta_0$ is the low-temperature expectation value of $\eta$.

In the short-time limit, the value of $\eta(t)$ changes extremely slow and the decay is dominated by a third-order term

$$\eta(t) = \eta_0 \frac{t^3}{3\tau_Q^2 \tau_\eta} + O(t^4)$$

which is much slower than $Q$. In the long-time limit (large $t$),

$$\eta(t) = \begin{cases} \eta_0 \left[ 1 - \dfrac{\tau_\eta^2}{(\tau_Q - \tau_\eta)\left(\frac{\tau_Q}{2} - \tau_\eta\right)} e^{-\frac{t}{\tau_\eta}} \right] & \tau_\eta > \tau_Q \\ \eta_0 \left( 1 - \dfrac{2\tau_Q}{\tau_Q - \tau_\eta} e^{-\frac{t}{\tau_Q}} \right) & \tau_\eta < \tau_Q \end{cases}$$

For $\tau_\eta > \tau_Q$, the change of $\eta$ is slower than $Q$. For $\tau_\eta < \tau_Q$, $\eta$ and $Q$ share the same decay time. In any of these two cases, *the variation of the secondary order parameter can never be faster than the primary one.*

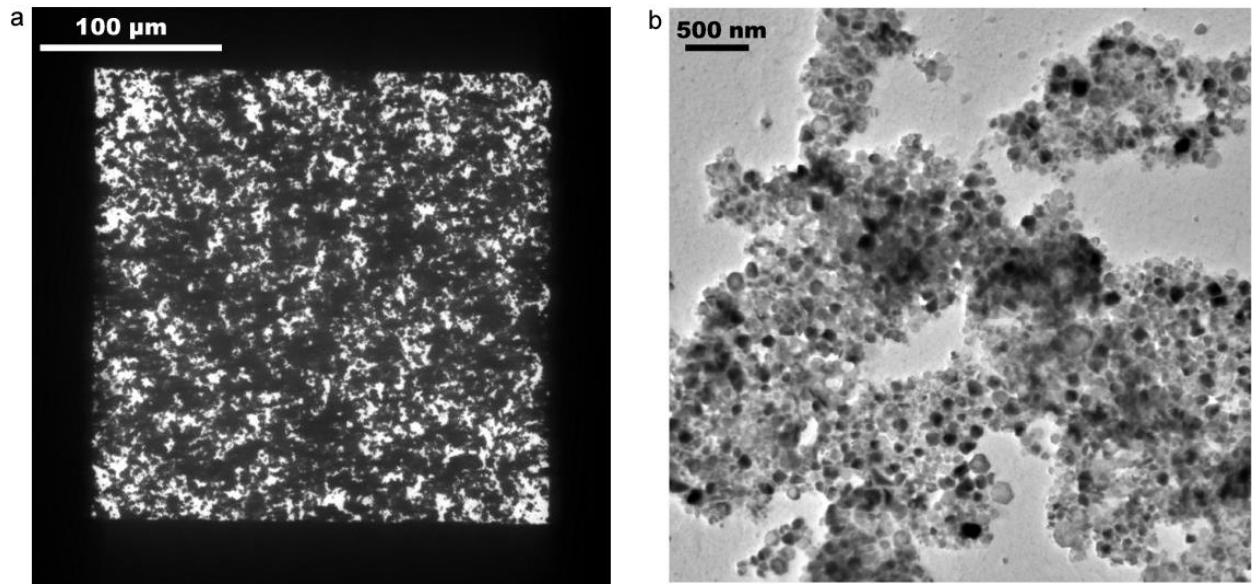

**Supplementary Figure S2.** (a) A low-mag TEM image of the $Cu_2S$ sample on a 15-nm-thick β-$Si_3N_4$ membrane TEM grid for the UED experiments. (b) A magnified TEM image obtained from a typical area of (a) shows most of the hexagonal $Cu_2S$ nanoplates laying down on β-$Si_3N_4$ support with their *c* axes aligned along the electron beam direction.

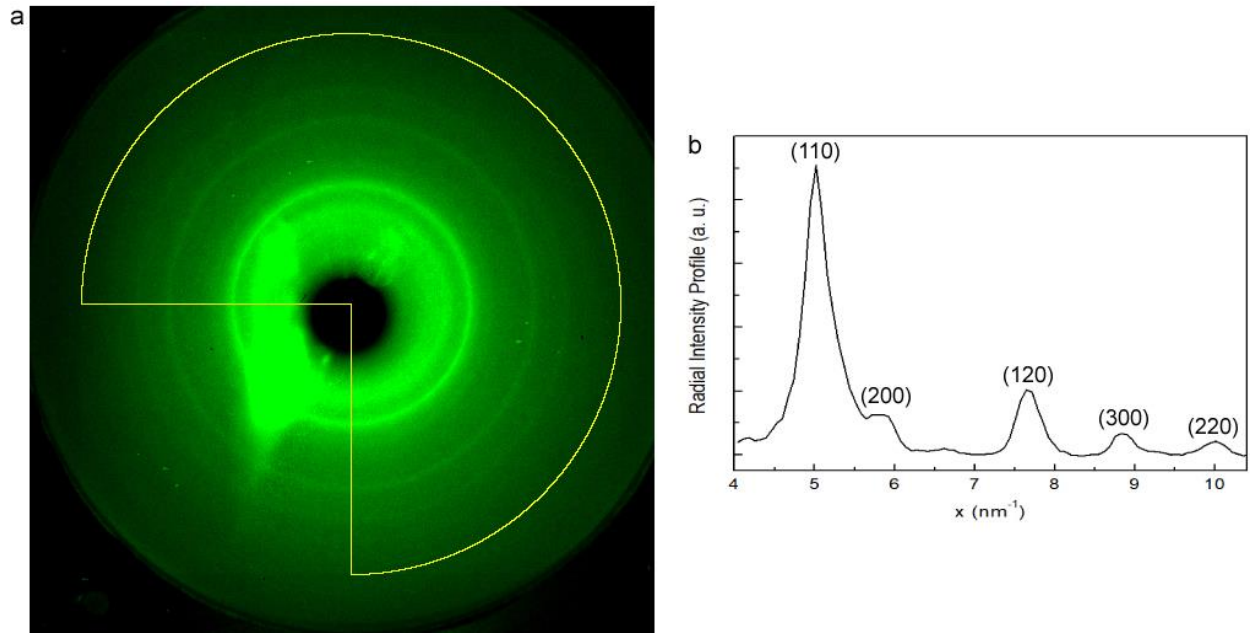

**Supplementary Figure S3.** (a) An UED pattern acquired at the time delay $t = 80$ ps after laser pumping. In order to avoid the part in the UED pattern that is polluted by the dark current, only three quarters of the pattern is used (highlighted by the yellow curve) for intensity analysis and measurements. The intensity profile of (a) is shown in (b) after a background subtraction using a curve fitting of combined Lorentzian (for small q) and exponential (for high q) functions. The indexed reflections are perpendicular to the *c* axis of the $Cu_2S$ nanoplates and confirm the TEM observation that most of the $Cu_2S$ nanoplates laying down on $\beta$-$Si_3N_4$ support.

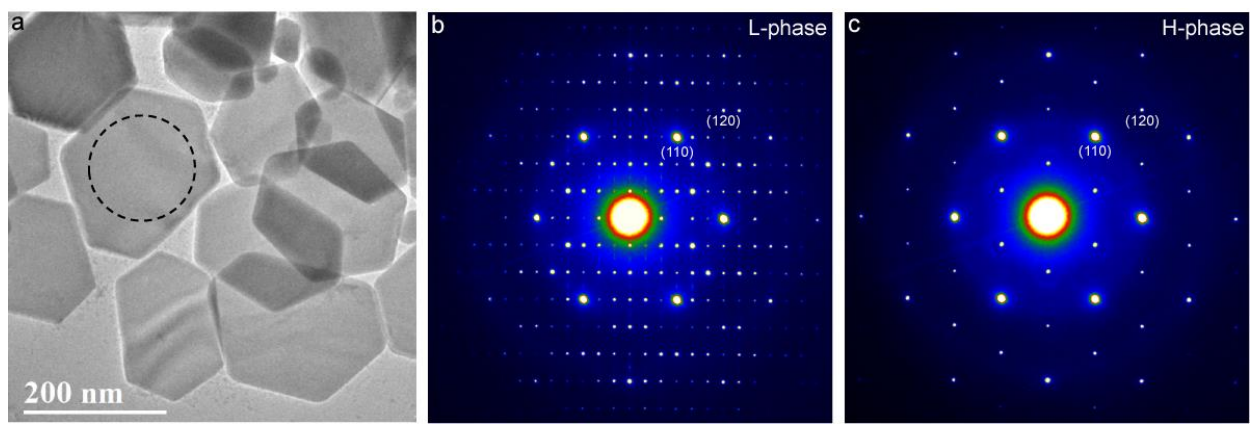

**Supplementary Figure S4.** (a) A TEM image shows individual $Cu_2S$ nanoplates and highlights the single-crystal area where the selected-area electron diffraction (ED) patterns were obtained using TEM techniques. (b) is from the monoclinic L-phase and (c) is from the hexagonal H-phase upon heating. The ED pattern from the H-phase (c) is purely hexagonal, whereas the ED pattern from the L-phase (b) shows superlattice reflections in addition to a set of nearly-hexagonal reflections.

| Phase \ Reflections | (110) | (120) |
|---|---|---|
| L-phase (at room temperature) | 2 reflections with single $q$ = 0.505 | 4 reflections with single $q$ = 0.771 |
| | | 4 reflections with single $q$ = 0.773 |
| | 4 reflections with single $q$ = 0.507 | 4 reflections with single $q$ = 0.774 |
| | | 4 reflections with single $q$ = 0.751 |
| | 4 reflections with single $q$ = 0.513 | 4 reflections with single $q$ = 0.804 |
| | | 4 reflections with single $q$ = 0.806 |
| H-phase (at 325 °C) | 6 reflections with single $q$ = 0.502 | 12 reflections with single $q$ = 0.767 |

Wave number (nm$^{-1}$)

**Supplementary Table S1.** This table provides the $q$ values of all the reflections that form the (110) and (120) peaks in the UED patterns shown in Fig. 2. The reflections in the table were measured from an individual single-crystal Cu$_2$S nanoplate using electron diffraction technique in a TEM (see supplementary Figure S4) and are consistent with the known structures [19, 21]. In the UED patterns, those reflections with close $q$ values are not distinguishable due to beam broadening and the UED peaks have the intensity profiles as the summation of the intensity profiles of listed reflections.

For the hexagonal H-phase, the (110) reflections in the ED pattern have single $q$ value of 5.02 nm$^{-1}$ and, similarly, a single $q$ value of 7.67 nm$^{-1}$ is related to the (120) reflections at 325 °C [21]. No additional reflections with nearby $q$ values appear around the (110) or (120) reflections in the ED pattern. Thus the (110) and (120) peaks in the UED pattern in Fig. 2(a) are formed by those single-$q$-valued reflections convoluted with instrument broadening.

On the other hand for the monoclinic L-phase, as highlighted by open-green squares (Fig. 2(b)), the (110) and (120) reflections have multiple q-values. The single-$q$-valued (110) reflection in the H-phase splits into multiple $q$ values of 5.05 nm$^{-1}$ and 5.07 nm$^{-1}$. Moreover, the ED pattern obtained from the L-phase has additional reflections due to its lower crystal symmetry, some highlighted by open-green triangles in Fig. 2(b). One of these additional reflections has a $q$ value of 5.13 nm$^{-1}$, very close to those of the (110) reflections. Consequently the (110) peak for the L-phase in the UED pattern (Fig. 2(a)) is formed as the summation of these three reflections (5.05 nm$^{-1}$, 5.07 nm$^{-1}$ and 5.13 nm$^{-1}$), giving an additional broadening of the peak intensity profile.

This is similarly true for the (120) reflections in the ED pattern (in open-green squares) which split into 7.71 nm$^{-1}$, 7.73 nm$^{-1}$ and 7.74 nm$^{-1}$ $q$ values in the L-phase. Furthermore there are additional reflections due to the lower symmetry (in open-green triangles) with nearby $q$ values that also contribute to the averaged intensity profile of the (120) peak in the UED pattern (Fig. 2(a)).